\documentclass[12pt,preprint]{aastex}
\usepackage{emulateapj5}
\usepackage{apjfonts}
\usepackage{natbib}

\input psfig.tex

\def\aa{{A\&A}}

\def\aj{{AJ}}
\def\al{$\alpha$}
\def\bet{$\beta$}

\def\annrev{{ARA\&A}}
\def\apj{{ApJ}}
\def\apjs{{ApJS}}
\def\asec{$^{\prime\prime}$}

\def\cc{cm$^{-3}$}
\def\deg{$^{\circ}$}

\def\cc{cm$^{-3}$}
\def\e#1{$\times$10$^{#1}$}
\def\etal{{et al. }}

\def\hst{{\it HST}}
\def\kms{km s$^{-1}$}
\def\lamb{$\lambda$}
\def\lax{{$\mathrel{\hbox{\rlap{\hbox{\lower4pt\hbox{$\sim$}}}\hbox{$<$}}}$}}
\def\gax{{$\mathrel{\hbox{\rlap{\hbox{\lower4pt\hbox{$\sim$}}}\hbox{$>$}}}$}}
\def\simlt{\lower.5ex\hbox{$\; \buildrel < \over \sim \;$}}
\def\simgt{\lower.5ex\hbox{$\; \buildrel > \over \sim \;$}}
\def\lum{erg s$^{-1}$}
\def\mbh{{$M_{\rm BH}$}}

\def\mnras{{MNRAS}}
\def\nat{{Nature}}
\def\pasp{{PASP}}

\def\percm2{cm$^{-2}$}
\def\peryr{yr$^{-1}$}

\def\solum{$L_\odot$}
\def\solmass{$M_\odot$}

\def\hii{\ion{H}{2}}

\def\nii{[\ion{N}{2}]}

\def\lhal{$L_{{\rm H}\alpha}$}
\def\lbol{$L_{{\rm bol}}$}
\def\ledd{$L_{{\rm Edd}}$}

\def\sigs{$\sigma_*$}
\def\mbh{{$M_{\rm BH}$}}

\slugcomment{To appear in {\it The Astrophysical Journal}.}
\shorttitle{ACCRETION IN NEARBY GALAXIES}
\shortauthors{HO}

\begin{document}

\title{Radiatively Inefficient Accretion in Nearby Galaxies}

\author{Luis C. Ho}

\affil{The Observatories of the Carnegie Institution of Washington, 813 Santa 
Barbara Street, Pasadena, CA 91101, USA}

\begin{abstract}
We use new central stellar velocity dispersions and nuclear X-ray and H\al\ 
luminosities for the Palomar survey of nearby galaxies to investigate the 
distribution of nuclear bolometric luminosities and Eddington ratios for their 
central black holes (BHs).  This information helps to constrain the nature of their 
accretion flows and the physical drivers that control the spectral diversity 
of nearby active galactic nuclei.  The characteristic values of the bolometric 
luminosities and Eddington ratios, which span over 7--8 orders of magnitude, 
from \lbol\ \lax\ $10^{37}$ to $3\times 10^{44}$ \lum\ and \lbol/\ledd\ 
$\approx 10^{-9}$ to $10^{-1}$, vary systematically with nuclear spectral 
classification, increasing along the sequence absorption-line nuclei 
$\rightarrow$ transition objects $\rightarrow$  low-ionization nuclear 
emission-line regions $\rightarrow$ Seyferts.
The Eddington ratio also increases from early-type to late-type galaxies.  We 
show that the very modest accretion rates inferred from the nuclear 
luminosities can be readily supplied through local mass loss from evolved 
stars and Bondi accretion of hot gas, without appealing to additional fueling
mechanisms such as angular momentum transport on larger scales.  Indeed, we 
argue that the fuel reservoir generated by local processes should produce far 
more active nuclei than is actually observed.  This generic luminosity-deficit
problem suggests that the radiative efficiency in these systems is much less 
than the canonical value of 0.1 for traditional optically thick, geometrically 
thin accretion disks.  The observed values of \lbol/\ledd, all substantially 
below unity, further support the hypothesis that massive BHs in most 
nearby galaxies reside in a low or quiescent state, sustained by accretion 
through a radiatively inefficient mode.
\end{abstract}


\section{Introduction}

Simple considerations of the quasar population predict that massive black 
holes (BHs) should be common in a sizable fraction of present-day galaxies.  
From the integrated luminosity density of quasars, one can estimate that 
a typical $L^*$ galaxy should, on average, contain a waste mass of
$\sim 10^7$ \solmass\ locked up in a BH (So\l tan 1982; Chokshi \& Turner 
1992).  But why are the quasar remnants so quiescent?  Active galactic nuclei 
(AGNs) with quasarlike luminosities are absent at $z\,\approx\,0$ presumably 
because of the diminished fuel supply currently available.  For a canonical 
radiative efficiency of $\eta$ = 0.1 appropriate for geometrically thin, 
optically thick accretion disks (see Frank et al. 1992), the BH has 
to consume 1--100 \solmass\ yr$^{-1}$ in order to generate luminosities of 
$L_{\rm bol}$ = 10$^{11}$--10$^{13}$ \solum, as typically seen in quasars.  
This level of gas supply is difficult to sustain in nearby galaxies.  
On the other hand, accretion rates of $\dot M$ = 0.001--0.1 \solmass\ 
yr$^{-1}$ do not appear implausible.  Even if angular momentum transport on 
nuclear scales is ineffective in disk galaxies, this level of fueling can
be supplied simply through local stellar mass loss (Ho et al.
1997c).  Hence, for $\eta$ = 0.1, there ought to be many nuclei shining as 
AGNs with $L_{\rm bol}$ = 10$^{9}$--10$^{11}$ \solum.  This is not observed.  
Only $\sim$1\%--5\% of galaxies contain bright Seyfert nuclei (e.g., Huchra \& 
Burg 1992; Greene \& Ho 2007a).  In the case of giant elliptical galaxies 
experiencing cooling flows, we may expect even larger values of $\dot M$ and 
thus correspondingly larger luminosities, again contrary to observations 
(Fabian \& Canizares 1988).

The dilemma posed by the luminosity deficit in the nuclei of nearby elliptical 
galaxies can be resolved by discarding the premise that massive BHs are 
ubiquitous in these systems (Fabian \& Canizares 1988).  But this proposition
is no longer tenable in light of our current knowledge on the demography 
of central BHs based on direct dynamical searches (Magorrian et al. 1998; Ho 
1999a; Kormendy 2004).  Massive BHs appear to be a generic component of 
galactic structure in most, if not all, systems with a bulge.  Consistent with 
this picture, low-level nuclear activity qualitatively resembling that of more 
luminous AGNs is found to be equally pervasive in nearby galaxies (Ho et al. 
1997b; Ho 2008).

Radiatively inefficient accretion flows (RIAFs; see Narayan et al. 1998;
Quataert 2001 for reviews) provide an attractive framework 
for solving the luminosity-deficit problem.   In the regime when the mass 
accretion rate onto the central BH is very low, the low-density, optically 
thin accreting medium cannot cool efficiently, and the accretion flow 
consequently puffs up into a quasi-spherical structure.  Most relevant to the 
present discussion, RIAFs attain radiative efficiencies much below the 
canonical value of 0.1.  RIAFs are characteristically dim.  Optically thin 
RIAFs are predicted to exist for accretion rates below a critical threshold of 
$\dot M_{\rm crit} \approx \alpha^2 \dot M_{\rm Edd} \approx 0.1 
\dot M_{\rm Edd}$ (Narayan et al. 1998), where the Eddington accretion rate 
is defined by $L_{\rm Edd}\,=\,\eta \dot M_{\rm Edd} c^2$, with $\eta \,=\,0.1$
and $L_{\rm Edd}\,=\,1.26 \times 10^{38} \left(M_{\rm BH}/M_{\odot}\right)$ 
\lum; the Shakura \& Sunyaev (1973) viscosity parameter is taken to be $\alpha 
\approx 0.3$ (Narayan et al. 1998).  Fabian \& Rees (1995) and Mahadevan 
(1997) invoke RIAFs to explain the dimness of elliptical galaxy nuclei.

This paper discusses the luminosity-deficit problem for a large, well-defined 
sample of galaxies spanning a wide range of morphological types and 
representing all the major nuclear spectroscopic classes.   We use X-ray and 
line luminosity measurements to constrain the accretion luminosities of the 
nuclei.  A newly published catalog of central stellar velocity dispersions 
provides estimates of the BH mass through the \mbh$-$\sigs\ relation (Gebhardt 
et al. 2000; Ferrarese \& Merritt 2000).  The feeble 

\begin{figure*}[t]
\hskip -0.3in
\psfig{file=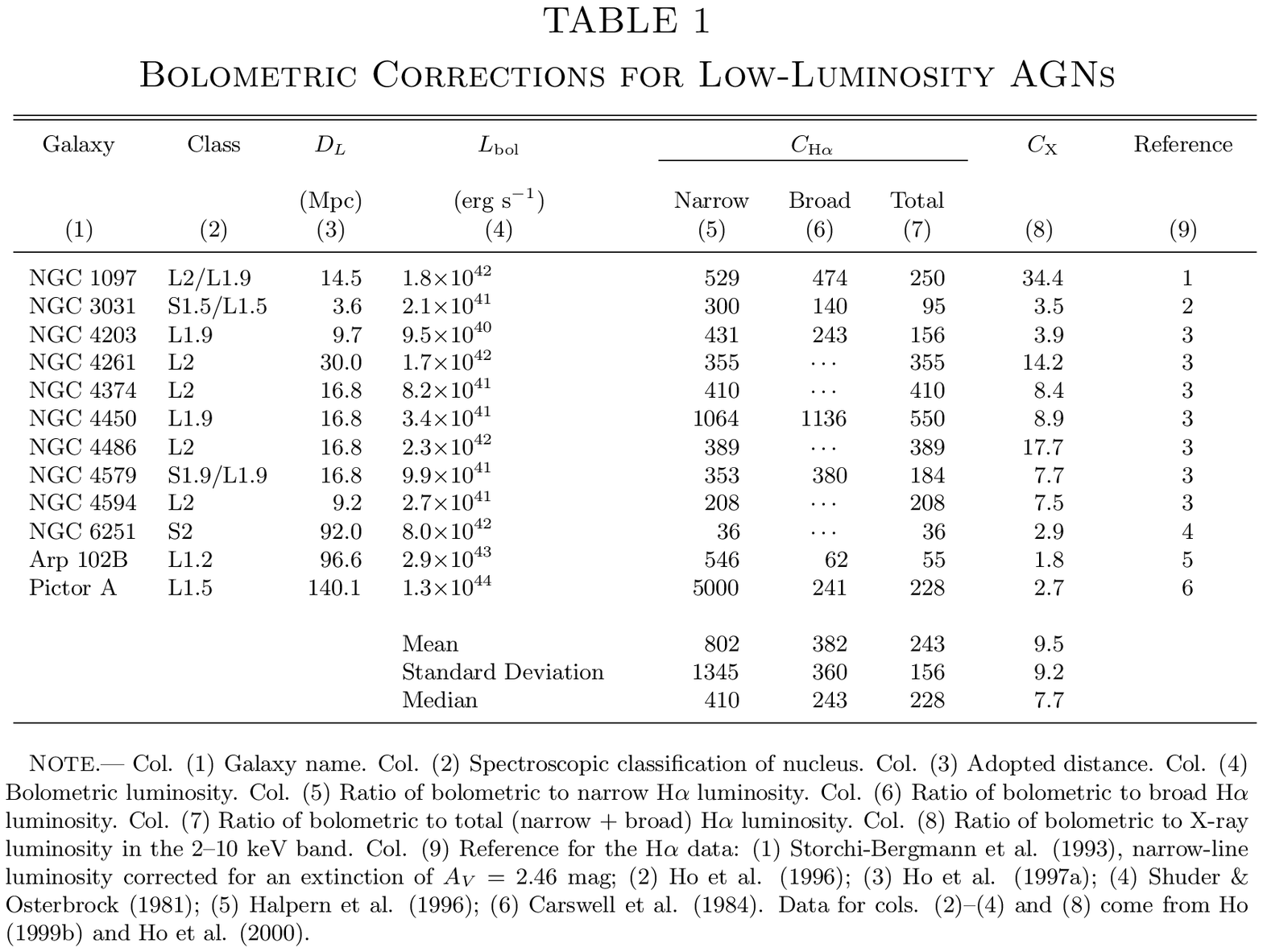,width=18.5cm,angle=0}
\end{figure*}

\noindent
nuclear activity in nearby
galaxies can be comfortably sustained through stellar mass loss and spherical 
accretion of hot gas in the inner regions of bulges.  We show that the 
inferred accretion rates lie well within the critical threshold of RIAFs.  
Finally, we discuss the physical connection between accretion rates and 
nuclear spectral types.

\section{Observational Material}

Our analysis explicitly makes the following assumptions: (1) all galaxy 
bulges contain central BHs whose masses can be well estimated using the 
recently established correlation between BH mass and bulge stellar velocity 
dispersion; (2) the present level of activity of the BHs manifests itself as 
AGN-like emission-line nuclei; and (3) the accretion luminosity can be 
extrapolated through the observed nuclear X-ray or optical line luminosities.

We focus on the sample of nuclei in the Palomar survey of nearby galaxies, a
magnitude-limited spectroscopic study of a nearly complete sample of 486
bright ($B_T \leq 12.5$ mag), northern ($\delta >$0\deg) galaxies (see Ho et
al. 1997a, 1997b, and references therein).  All of the galaxies have been 
assigned a nuclear spectroscopic classification using a set of uniform 
criteria, into the following classes (see Ho et al. 1997a for details):
absorption-line nuclei (A), \hii\ nuclei (H), Seyfert nuclei (S), 
low-ionization nuclear emission-line regions (LINERs; L),
and transition objects (T; LINER/\hii\ composites).  This paper considers all
classes except the \hii\ nuclei.  The Palomar survey contains 277 galaxies 
classified as absorption-line (66), Seyfert (52), LINER (94), and transition 
nuclei (65).

\subsection{Nuclear Luminosities}

Our study requires bolometric luminosities for a well-defined sample of AGNs 
covering a wide range of power.  Because AGNs emit a very broad spectrum, 
their bolometric luminosities ideally should be measured directly from their 
broadband spectral energy distributions (SEDs).  In practice, however, 
complete SEDs are not readily available for most AGNs, least of all for 
low-luminosity sources such as LINERs, which are most prevalent in nearby 
galaxies.  The largest existing compilations of broadband SEDs for 
low-luminosity AGNs (Ho 1999b; Ho et al. 2000; Maoz 2007) contain only a 
limited number of objects.  

For this study, we circumvent this difficulty by using two measures of the 
nuclear power to estimate the AGN bolometric luminosity, one based on the 
H\al\ emission line and another based on X-rays.  Although the H\al\ luminosity 
comprises only a small percentage of the total power, its fractional 
contribution to the bolometric luminosity, as shown below, turns out to be 
fairly well defined.  Moreover, unlike most other spectral windows, H\al\ 
measurements are readily available for large, relatively complete samples of 
nuclei.  The Palomar database gives H\al\ luminosities measured through an 
aperture of 2\asec$\times$4\asec\ centered on the nucleus, which corresponds 
to a linear scale of $\sim$200 pc $\times$ 400 pc for a typical distance of 20 
Mpc.  As explained in Ho et al. (2003a), some of the H\al\ luminosities 
published in Ho et al. (1997a) have been updated with more accurate values 
from the literature.  We will also make use of upper limits for the H\al\ 
luminosity of the absorption-line nuclei that were not given explicitly in 
Ho et al. (1997a); the limits were calculated from the equivalent-width 
detection limit of the survey in conjunction with the 6600 \AA\ continuum flux 
density measurements, assuming that the line has a typical  full width at 
half-maximum (FWHM) of 250 \kms.  A supplementary list of H\al\ luminosities, 
including the upper limits, is given in Ho et al. (2003a).  In total, H\al\ 
luminosities, or upper limits thereof, are available for 246 objects, which 
account for 80\%, 98\%, 89\%, and 91\% of the nuclear classes A, S, L, and T, 
respectively.  Thus, H\al\ measurements are available for the vast majority of 
the objects considered in this study.

The bolometric correction for H\al, $C_{\rm H\alpha} = L_{\rm{bol}}/
L_{\rm H\alpha}$, can be obtained in one of two ways.  For luminous, type~1 
AGNs, it is often expedient to estimate $L_{\rm{bol}}$ from the optical 
continuum luminosity, frequently chosen at 5100 \AA\ in the recent literature: 
$L_{\rm{bol}} = C_{\rm 5100 \AA} \lambda L_\lambda({\rm 5100 \AA})$.  Now, the 
H\al\ luminosity correlates strongly with the optical continuum, including 
$\lambda L_\lambda({\rm 5100 \AA})$ (Greene \& Ho 2005); the correlation 
is slightly nonlinear.  Choosing $C_{\rm 5100 \AA} = 9.8$ (McLure \& Dunlop 
2004), Greene \& Ho (2007a) obtain $L_{\rm{bol}} = 2.34 \times 10^{44} 
(L_{\rm H\alpha} / 10^{42}\,{\rm erg~s^{-1}})^{0.86}$ \lum.  The conversion 
pertains to the entire H\al\ line, which in luminous type~1 AGNs is dominated 
by the broad component.  Because of the reliance on the \lhal$-$$\lambda 
L_\lambda({\rm 5100 \AA})$ correlation and the assumption that $C_{\rm 5100 
\AA}$ is constant, the relation between \lbol\ and \lhal\ is formally slightly 
nonlinear.  It is unclear how robust this result is and whether it can be 
extrapolated toward lower luminosities.  For the luminosity range of interest 
to us, it introduces an uncertainty of a factor of $\sim 2$ into the 
bolometric correction. For \lhal\ = $10^{38}$, $10^{39}$, and $10^{40}$ \lum, 
$C_{\rm H\alpha} \approx 850$, 615, and 446, respectively.

Alternatively, we can attempt to estimate $C_{\rm H\alpha}$ empirically from 
the observed SEDs of low-luminosity AGNs, limited though they may be.  We use 
the sample of 12 low-luminosity Seyferts and LINERs with broadband SEDs 
studied by Ho (1999b) and Ho et al. (2000; see also Ho 2002b) as a guide.  The 
data, summarized in Table~1, show that the median value of $C_{\rm H\alpha}$
ranges from 228 to 410, depending on whether we include only the narrow 
component of the line, only the broad component, or both.  A reasonable 
compromise might be $C_{\rm H\alpha} \approx 300\pm100$.  Preliminary analysis 
of a more extensive sample of SEDs (L. C. Ho, in preparation) shows that the 
bolometric correction for the H\bet\ line has a significant scatter, especially
for low-luminosity sources.  For sources with Eddington ratios below 0.1, and 
assuming that on average H\al/H\bet = 3.5 (Greene \& Ho 2005), the median 
value of $C_{\rm H\alpha} \approx 220$ with an interquartile range of 160. 
This agrees reasonably well with the range of values obtained from our 
calibration sample in Table~1.   For concreteness, we will simply adopt 
$C_{\rm H\alpha} = 300$; none of the main conclusions in this study depends
critically on the exact value of the bolometric correction.

The H\al\ luminosities of the narrow-line objects are subject to a potential
source of complication.  From consideration of the relative strength of X-ray 
(2--10 keV) and H\al\ emission in different classes of nearby AGNs, Ho (2008) 
shows that type~2 sources (Seyfert 2s, LINER 2s, and essentially all of the 
transition objects) have a general tendency to emit excess optical line 
emission compared to their type~1 counterparts.  He attributes this excess 
emission to extranuclear processes unassociated with the active nucleus.  If 
this interpretation is correct, then only a fraction of the narrow H\al\ 
emission should be included in the budget for the nuclear luminosity.   Ho 
(2008) finds that the median ratio of $L_{\rm X}/L_{\rm H\alpha}$ is 7.3 for 
Seyfert~1s and 4.6 for LINER~1s, whereas Seyfert~2s, LINER~2s, and transition 
objects have significantly lower values of 0.75, 1.6, and 0.41, respectively.
Assuming, for concreteness, that the intrinsic $L_{\rm X}/L_{\rm H\alpha}$ 
ratio for the type~2 sources is equal to that of LINER~1s, the H\al\ 
luminosities for Seyfert~2s, LINER~2s, and transition objects need to be 
reduced by a factor of 6.1, 2.9, and 11.2, respectively.

In light of the above source of ambiguity with the H\al\ emission, it would be 
desirable to have an alternative handle on the nuclear luminosity.  By far the 
most secure measure of nuclear luminosity in AGNs comes from X-ray observations.
Because of the faintness of low-luminosity AGNs, the data must be of 
high enough angular resolution so that the nucleus can be cleanly separated 
from the host galaxy (e.g., Ho et al. 2001; Flohic et al. 2006).  As reviewed 
in Ho (2008), a substantial fraction of the galaxies in the Palomar survey 
have now been observed in the X-rays.  The Appendix gives a compilation of all 
pertinent X-ray measurements taken from the literature, as well as from new 
analysis of data taken from the {\it Chandra}\ public archives.  The X-ray 
data are not nearly as complete as H\al.  Nevertheless, X-ray luminosities, or 
upper limits thereof, are now available for 175 out of the 277 objects in the 
parent sample (63\%), which account for 47\% of the absorption nuclei, 68\% of 
the LINERs, 83\% of the Seyferts, and 57\% of the transition objects.  The 
incompleteness and heterogeneous nature of the X-ray measurements make it 
difficult to rigorously assess selection effects.  However, if observational 
biases exist, they should be in the direction of missing very faint sources, 
an effect that strengthens our main conclusions.

As with the H\al\ data, determining the appropriate bolometric correction for
the X-ray band is not trivial.  Since the SEDs of AGNs vary strongly with
accretion rate (Ho 1999b), we must abandon the usual practice of adopting a
single correction factor based on the average SED of luminous quasars.  Using,
again, the small sample of low-luminosity AGNs with reliable broadband SEDs,
Table~1 shows that the median value of the bolometric correction in the 2--10
keV band is $C_{\rm X} = L_{\rm{bol}}/L_{\rm X} \approx 8$, where 
$L_{\rm X}$ is the luminosity in the 2--10 keV band, corrected, to the
extent possible, for absorption.  The more extensive data set of L. C. Ho (in
preparation) suggests a value larger by about a factor of 2: sources with
\lbol/\ledd\ \lax\ 0.1 have a median $C_{\rm X} = 15.8$, with an interquartile 
range of 9.6.  Because low-luminosity AGNs tend to be ``X-ray-loud'' (Ho 1999b) their 
values of $C_{\rm X}$ are significantly smaller than 
conventionally assumed for luminous sources ($C_{\rm X} \approx 35$; Elvis
et al. 1994).  This is consistent with the analysis of Vasudevan \& Fabian
(2007).  For the present purposes, we will adopt $C_{\rm X} = 15.8$, noting, as
before, that factors of a few variation in the bolometric correction do not 
affect the main conclusions of this study.

The uncertainties on \lbol\ are difficult to estimate.  As discussed in Ho et 
al. (1997a), the H\al\ fluxes in the Palomar survey have typical errors of 
30\%--50\%, reaching 100\% in the worst 

\vskip 0.3cm
\begin{figure*}[t]
\centerline{\psfig{file=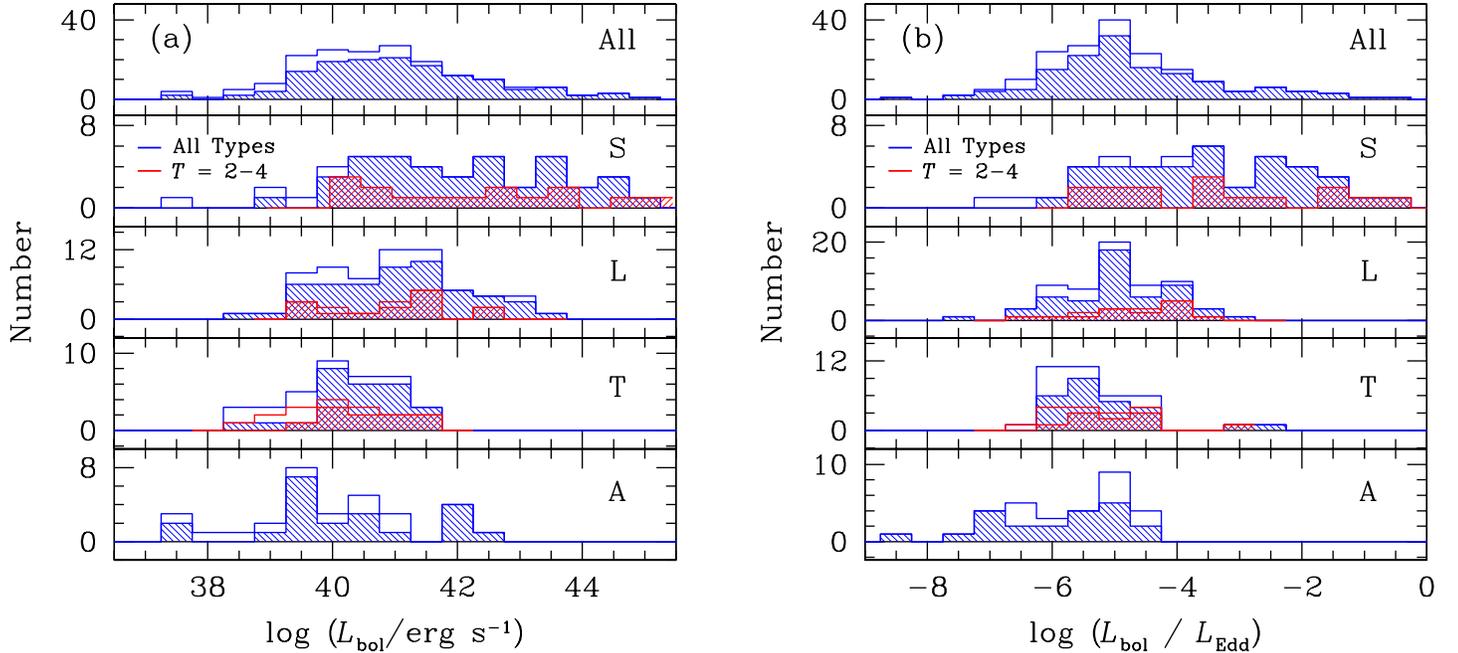,width=19.5cm,angle=0}}
\figcaption[fig1.ps]{
Distribution of ({\it a}) bolometric luminosity, $L_{\rm bol}$, and ({\it b})
ratio of bolometric luminosity to the Eddington luminosity,
$L_{\rm bol}/L_{\rm Edd}$, for all objects, Seyferts (L), LINERs (L),
transition nuclei (T), and absorption-line nuclei (A).  The hatched and
open histograms denote detections and upper limits, respectively.  The
original sample is shown in blue, and the subsample restricted to Sab--Sbc
($T = 2-4$) Hubble types is shown in red.  The bolometric luminosity is based
on the 2--10 keV X-ray luminosity, assuming $L_{\rm bol} = 15.8 L_{\rm X}$.
\label{fig1}}
\end{figure*}
\vskip 0.3cm

\vskip 0.3cm
\begin{figure*}[t]
\centerline{\psfig{file=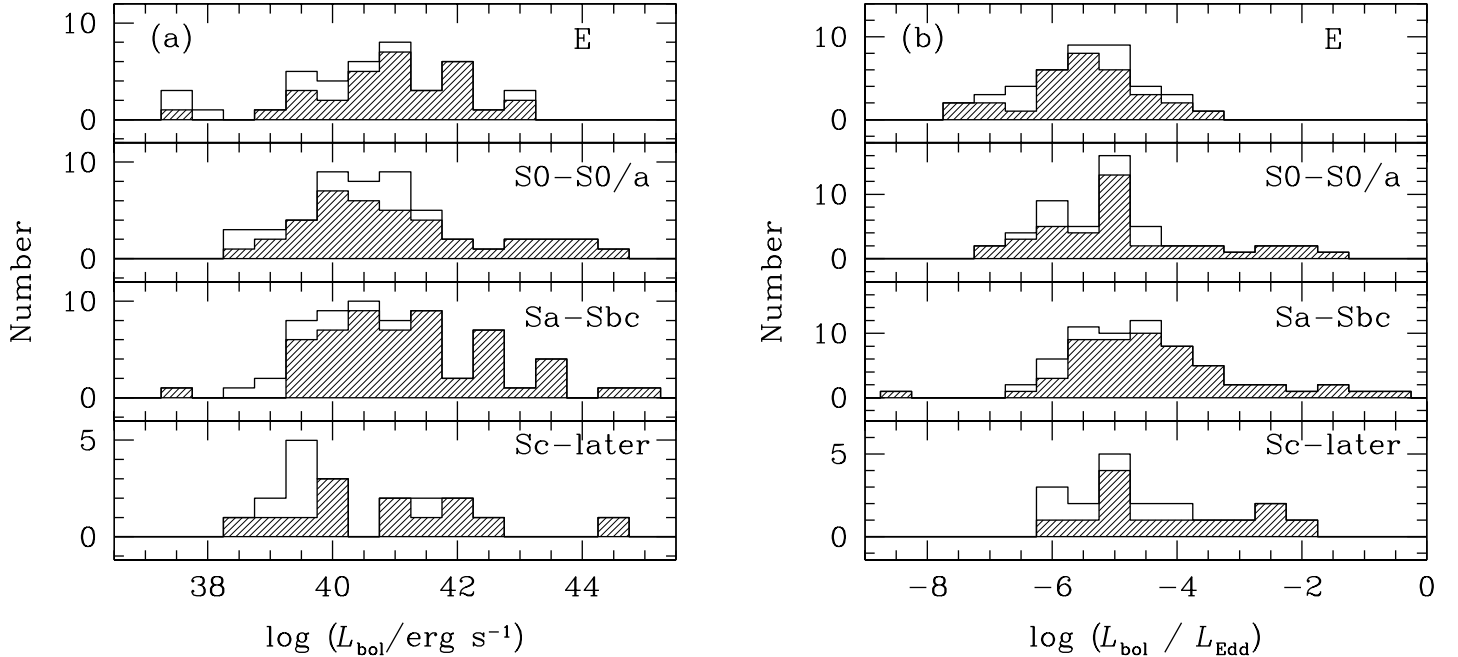,width=19.5cm,angle=0}}
\figcaption[fig2.ps]{
Distribution of ({\it a}) bolometric luminosity, $L_{\rm bol}$, and ({\it b})
ratio of bolometric luminosity to the Eddington luminosity,
$L_{\rm bol}/L_{\rm Edd}$, for galaxies binned by Hubble type.  The hatched
and open histograms denote detections and upper limits, respectively.  The
bolometric luminosity is based on the 2--10 keV X-ray luminosity, assuming
$L_{\rm bol} = 15.8 L_{\rm X}$.
\label{fig2}}
\end{figure*}
\vskip 0.3cm

\noindent
cases.  The largest source of 
uncertainty for the H\al-based luminosities, however, comes from our 
still-tentative knowledge of $C_{\rm H\alpha}$ (factor $\sim 2$) and the 
amount of extranuclear contamination of the narrow-line emission (factor
$\sim 2-5$, depending on the spectral class).  In the X-ray band, we can be
more confident that the flux is largely nuclear, and large-amplitude
variability seems to be rather uncommon for the systems in question (Ho 2008).
Not all of the X-ray detections have sufficient counts for rigorous spectral
fitting, but fortunately low-luminosity AGNs generally have small absorbing
columns (Ho 2008).  Still, at the moment we do not know $C_{\rm X}$ to better
than a factor of $\sim 2$.  We conservatively guess that the estimates of
\lbol\ based on H\al\

\vskip 0.3cm
\begin{figure*}[t]
\centerline{\psfig{file=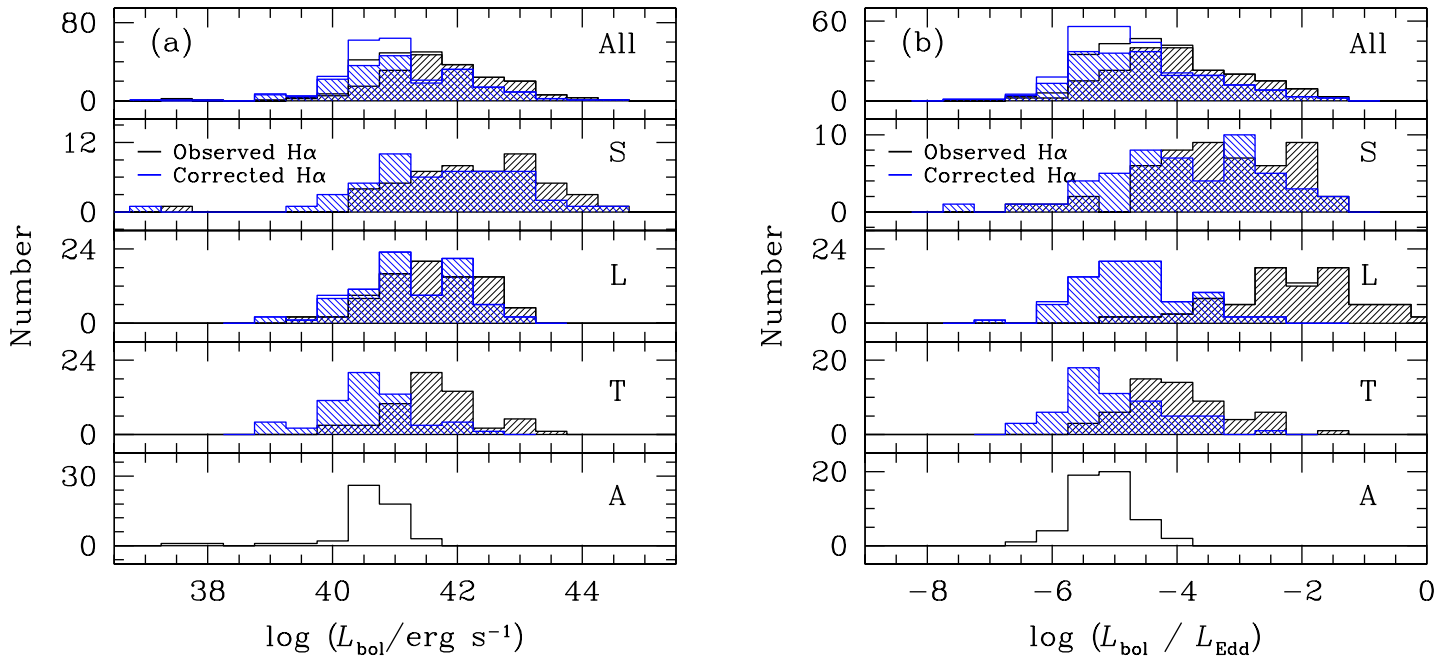,width=19.5cm,angle=0}}
\figcaption[fig3.ps]{
Distribution of ({\it a}) bolometric luminosity, $L_{\rm bol}$, and ({\it b})
ratio of bolometric luminosity to the Eddington luminosity,
$L_{\rm bol}/L_{\rm Edd}$, for all objects, Seyferts (L), LINERs (L),
transition nuclei (T), and absorption-line nuclei (A).  The bolometric
luminosity is based on the extinction corrected total (narrow + broad) H\al\
luminosity, assuming $L_{\rm bol} = 300 L_{\rm H\alpha}$.  The blue histograms
show the distributions after correcting the narrow-line sources for
extranuclear contamination (see text for details).  The hatched and open
histograms denote detections and upper limits, respectively.
\label{fig3}}
\end{figure*}
\vskip 0.3cm

\noindent
and X-rays measurements have uncertainties of 0.7 and
0.3 dex, respectively.  

\subsection{Black Hole Masses}

The majority of the objects in our sample do not have direct, dynamically
determined BH masses.  We will estimate BH masses using the tight correlation
between BH mass and bulge stellar velocity dispersion (the \mbh$-$$\sigma_*$ 
relation: Gebhardt et al. 2000; Ferrarese \& Merritt 2000), as determined by 
Tremaine et al. (2002):

\begin{equation}
\log\,\left(\frac{M_{\rm BH}}{M_{\odot}}\right)\, = \,
(4.02\pm0.32)\, \log\,\left(\frac{\sigma_*}{200\, {\rm km\,s}^{-1}}\right)\, 
+ \, (8.13\pm0.06).
\end{equation}

\vspace{0.3cm}
\noindent
The intrinsic scatter of the above fit is estimated to be \lax 0.3 dex.  The 
fit uses $\sigma_e$, the luminosity-weighted velocity dispersion measured 
within the effective radius of the bulge.  Since we do not have measurements of 
$\sigma_e$ for most of our galaxies, we use instead $\sigma_0$, the central 
velocity dispersion.  Gebhardt et al. (2000) have shown that in general 
$\sigma_0 \approx \sigma_e$ within a scatter of $\sim$10\%.

Ho et al. (2009) recently published a comprehensive, uniform catalog of central 
velocity dispersions for nearly all of the galaxies in the Palomar survey.  
New stellar velocity dispersions were obtained for a total of 428 galaxies, 
and estimates for another 34 were obtained indirectly from the line width of 
\nii\ \lamb6583 using the calibration of Ho (2009).  The typical 
uncertainties in the velocity dispersions range from $\sim$5\% to 15\%. An 
error of 10\% in $\sigma_*$ introduces an uncertainty of $\sim$0.15 dex in 
$\log$~\mbh.  We assume that \mbh\ has an uncertainty dominated by the 
scatter of the \mbh$-$$\sigma_*$ relation, $\sim$0.3 dex.

\vskip 1.0cm

\section{Bolometric Luminosities and Eddington Ratios}

Figure~1 shows the distributions of bolometric luminosities and their values 
normalized with respect to the Eddington luminosity, \lbol/\ledd.  To avoid
the possible complication of extranuclear contamination in the H\al\ emission,
we base the bolometric luminosities on the hard X-ray measurements.  The 
statistics of the distributions are listed in Table~2.   The four classes of 
nuclei comprise a sequence of increasing luminosity: A $\rightarrow$ T 
$\rightarrow$ L $\rightarrow$ S.  Whereas LINER and transition nuclei have 
very similar H\al\ luminosities (Ho et al. 2003a)---an effect that can be 
attributed to the H\al\ emission in transition objects being boosted by 
nonnuclear sources (Ho 2008)---there is no doubt that in the hard X-ray 
band LINERs are more luminous than transition objects (median \lbol\ = $3.0 
\times 10^{40}$ vs. $6.5 \times 10^{39}$ \lum).  Both LINERs and transition 
objects, in turn, are less powerful than Seyferts (median \lbol\ = $2.2 \times 
10^{41}$ \lum).  This systematic trend becomes even more sharply defined when 
we consider the Eddington ratios.  As with \lbol, the median value of 
\lbol/\ledd\ for LINERs ($6.0\times 10^{-6}$) is a factor of 4 larger than 
for transition objects ($1.5\times 10^{-6}$), both being 1--2 orders 
of magnitude smaller than in Seyferts ($1.1\times 10^{-4}$).  Notably, 
{\it all}\ galactic nuclei in the Palomar sample are highly sub-Eddington 
systems.

The distribution of \lbol\ is broadly similar for galaxies of different 
morphological types (Fig.~2{\it a}).  By contrast, \lbol/\ledd\ increases 
mildly, but systematically, from early-type to late-type galaxies 
(Fig.~2{\it b}).  Since the various classes of emission-line nuclei in the 
Palomar survey are hosted by slightly different Hubble types (Ho et al. 2003a), 
it would be of interest to examine the trends in \lbol\ and \lbol/\ledd\ after 
factoring out the dependence on Hubble type.  This is illustrated by the red 
histograms in Figure~1, where we now restrict the comparison to galaxies with 
morphological types Sab--Sbc ($T = 2-4$), a subset that, as shown in Ho et al. 
(2003a), has statistically identical 

\begin{figure*}[t]
\psfig{file=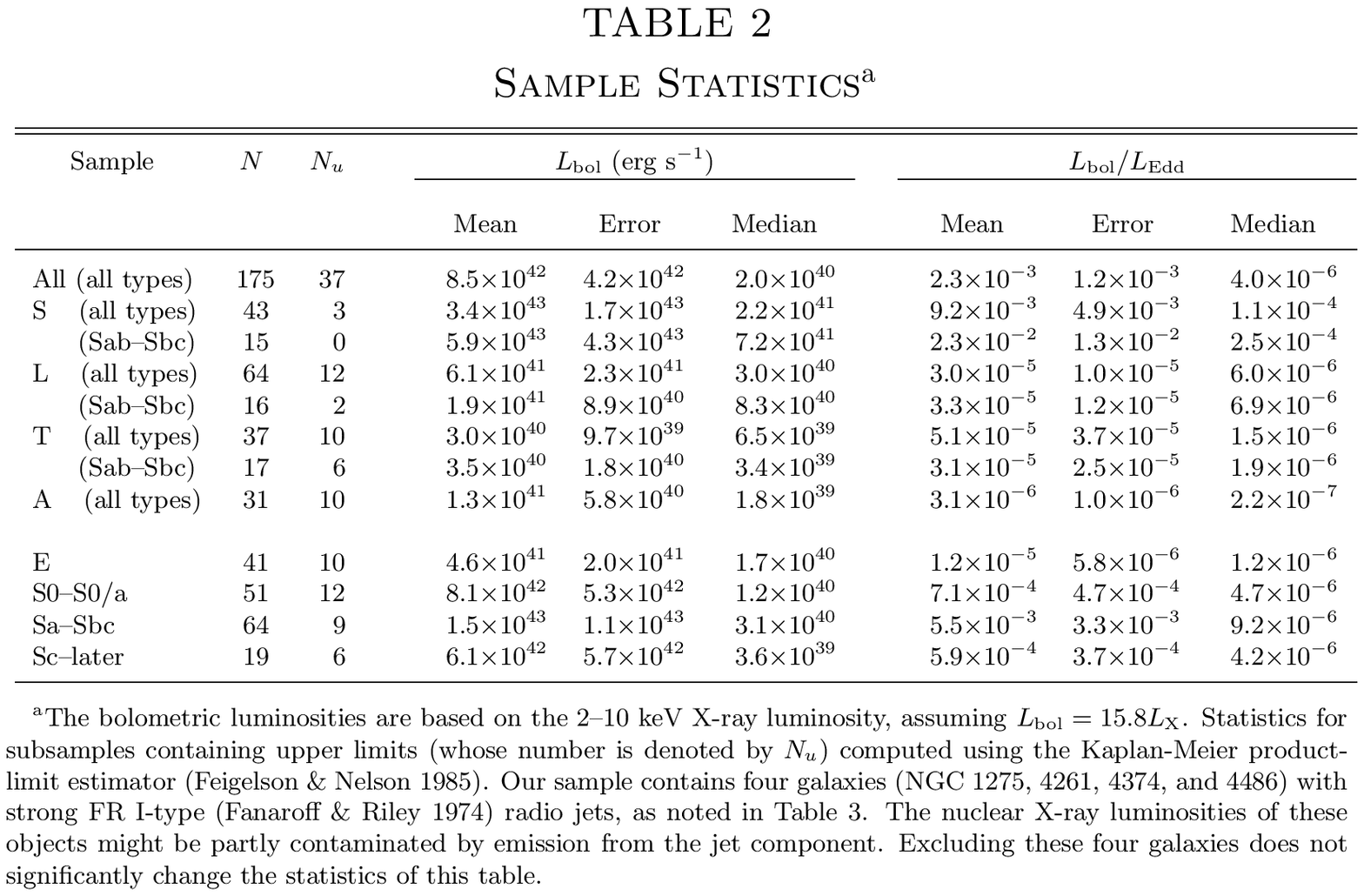,width=18.5cm,angle=0}
\end{figure*}

\noindent
distributions of morphological types for 
all three AGN subclasses.  The overall trends of the parent sample are 
preserved (see also Table~2).

Although in this study we give preference to the X-ray luminosities over the
H\al\ luminosities because of concerns over extranuclear line contamination,
the H\al\ data have the advantage of being uniform and nearly complete.
Figure~3 repeats the analysis of Figure~1, but now using bolometric 
luminosities derived from H\al.  The black histograms show the H\al\ 
luminosities as observed, converted to \lbol\ assuming a bolometric correction 
of $C_{\rm H\alpha} = 300$.  The blue histograms plot the same data with a 
statistical correction for extranuclear line emission applied to the 
Seyfert~2s, LINER~2s, and transition objects based on their observed 
$L_{\rm X}/L_{\rm H\alpha}$ ratio (see \S~2.1).  Not surprisingly, the 
overall trends seen in Figure~1 are well mirrored in Figure~3, but now they are
delineated with better statistics.

Finally, we turn to the distribution of \mbh\ and \lbol\ vs. \lbol/\ledd\
(Fig.~4).  In these diagrams, we have marked the various subclasses of nuclei, 
and we have included the large sample of $z < 0.35$ high-luminosity AGNs 
(Seyfert 1 nuclei and low-redshift quasars) selected by Greene \& Ho (2007a) 
from the Sloan Digital Sky Survey (SDSS).  Greene \& Ho derived BH masses and 
bolometric luminosities using the width and strength of the broad H\al\ line.
We point out several salient features.

\begin{enumerate}

\item{Considered collectively, there is no dependence of \mbh\ on \lbol/\ledd: 
at a given value of \mbh, which mostly fall in the range $10^6 - 10^9$ 
\solmass, \lbol/\ledd\ spans $\sim 6$ orders of magnitude within the Palomar
sample and $\sim 8$ orders of magnitude if the sample of luminous sources 
is included.}

\item{Within each class of emission-line objects, especially among LINERs and 
transition nuclei, there is a loose inverse correlation between \mbh\ and 
\lbol/\ledd.  This arises because \lbol\ spans a narrower range of values than 
\mbh.}

\item{At a fixed value of \mbh, \lbol/\ledd\ increases systematically along the 
sequence absorption-line nuclei $\rightarrow$ transition objects $\rightarrow$ 
LINERs $\rightarrow$ low-luminosity Seyferts $\rightarrow$ high-luminosity 
Seyferts and quasars. There is considerable overlap among the classes.  The 
apparent gap in \lbol/\ledd\ between the Palomar and SDSS sample may be an 
artifact of observational selection effects; the two surveys have very 
different sensitivity limits (Ho 2008).}

\item{All emission-line nuclei in nearby galaxies are sub-Eddington systems, 
with the vast majority having \lbol/\ledd\ $\ll$ 1.  All LINERs and transition 
nuclei are characterized by \lbol/\ledd\ $<\,10^{-2}$.}

\item{The combined distribution of \lbol\ or \lbol/\ledd\ for the Palomar 
sample shows no evidence for bimodality or other indications of an abrupt 
transition between low-ionization (LINERs and transition objects) and 
high-ionization (Seyferts) sources.}

\item{Because \lbol\ spans a much larger range than \mbh, \lbol\ broadly 
increases with increasing \lbol/\ledd.  Low luminosity generally corresponds 
to low Eddington ratios.  But there are important exceptions.  A minority of 
AGNs have low luminosities because they have low BH masses, not necessarily 
low Eddington ratios.  Within the Palomar sample, NGC~4395 (Filippenko \& Ho 
2003) provides a good example (Fig.~4{\it b}), and similar types of low-mass 
AGNs have been discovered in SDSS (Greene \& Ho 2004, 2007b).}
\end{enumerate}

\vskip 0.3cm
\begin{figure*}[t]
\centerline{\psfig{file=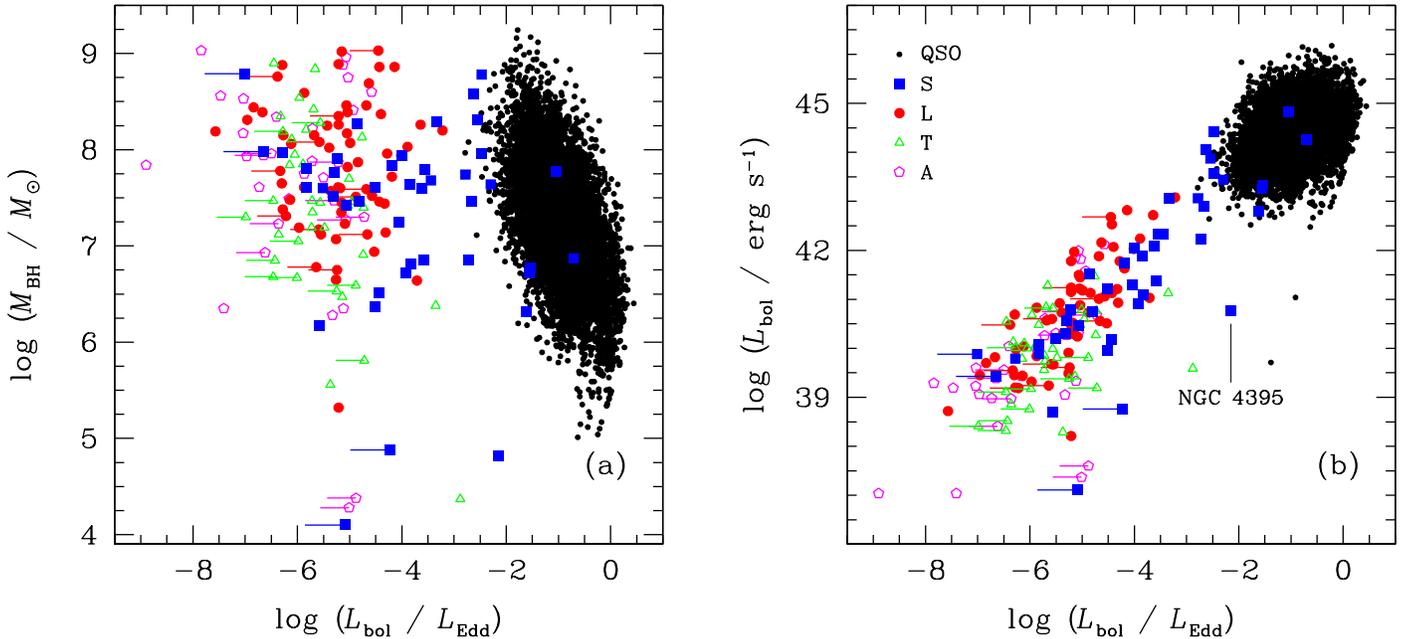,width=19.0cm,angle=0}}
\figcaption[fig4.ps]{
Distribution of ({\it a}) BH masses and ({\it b}) \lbol\ vs.
$L_{\rm bol}/L_{\rm Edd}$ for objects separated by spectral classification.
The bolometric luminosity is based on the 2--10 keV X-ray luminosity, assuming
$L_{\rm bol} = 15.8 L_{\rm X}$.  The symbols are identified in the legend.
The objects marked as ``QSO'' refer to the sample of high-luminosity Seyfert 1
nuclei and quasars studied by Greene \& Ho (2007a).  Line segments
denote upper limits.
\label{fig4}}
\end{figure*}
\vskip 0.3cm

\section{Sources of Fuel}

In this section we make some rough estimates of the {\it minimum}\ amount of
fuel likely to be available in the central regions of nearby galaxies.  For
the moment, let us neglect any contribution due to dissipation from a 
large-scale disk, external acquisition from tidal interactions and infall, or 
to discrete, episodic events such as tidal disruptions of stars.  Galactic 
nuclei can be fed, in a steady state manner from the inner bulge of the galaxy, 
by at least two sources: (1) ordinary mass loss from evolved stars and (2) 
gravitational capture of gas from the hot interstellar medium.

\subsection{Stellar Mass Loss}

Present-day elliptical galaxies and the bulges of S0s and spirals contain 
mostly old, evolved stars.  Red giants and planetary nebulae return a 
significant fraction of their mass to the interstellar medium through mass 
loss.  For a Salpeter stellar initial mass function with a lower mass cutoff 
of 0.1 \solmass, an upper mass cutoff of 100 \solmass, solar metallicities, 
and an age of 15 Gyr, Padovani \& Matteucci (1993) estimate 

\begin{equation}
\dot M_*\,\approx\,3\times10^{-11} \, \left(\frac{L}{L_{\odot, V}}\right)
\,\,\,\,\,\,\, M_{\odot}\,{\rm yr}^{-1}.
\end{equation}

\vspace{0.3cm}
\noindent
This result is consistent, within a factor of $\sim$2, with the work of Faber 
\& Gallagher (1976), Ciotti et al. (1991), and Jungwiert et al.
(2001).  Athey et al. (2002) obtained mid-infrared observations to 
probe more directly the mass-losing stars in elliptical galaxies.  They find 
$\dot M_*\,\approx\,7\times10^{-12} \left(L/L_{\odot, B}\right)$ \solmass\ 
\peryr.  For $B-V \approx 1$ mag, appropriate for an evolved population (e.g., 
Fukugita et al. 1995), $\dot M_*\,\approx\,2\times10^{-11} \, 
\left(L/L_{\odot, V}\right)$ \solmass\ \peryr, again close to the estimate by 
Padovani \& Matteucci (1993).  Thus, we use Padovani \& Matteucci's relation 
to convert between $V$-band luminosity and mass loss rate.

{\it Hubble Space Telescope (HST)}\ images have offered an unprecedently 
detailed view of the morphological structure of the central regions of nearby 
galaxies.  The majority of galaxies contain central density concentrations, 
either in the form of nuclear cusps or photometrically distinct, compact 
stellar nuclei (e.g., Lauer et al. 1995; Phillips et al. 1996; Carollo et al. 
1997; Faber et al. 1997; Rest et al. 2001; Ravindranath et al. 2001; B\"oker 
et al. 2002; Ferrarese et al. 2006; Kormendy et al. 2009).   The cusp profiles 
continue to rise to the resolution limit of \hst\ (0\farcs1), which is 
$\sim$10 pc at a distance of 20 Mpc.  The nuclear stellar population in 
most instances is old (Ho et al. 2003a; Sarzi et al. 2005; Zhang et 
al. 2008).  How much gaseous material is available through stellar mass loss?  
We note that the uncertainties associated with the effectiveness of angular 
momentum transport on large (1--10 kpc) or intermediate (0.1--1 kpc) scales 
are bypassed by focusing only on {\it nuclear}\ (\lax 10 pc) scales.  Although 
the fate of the nuclear gas is not entirely clear, it is important to 
recognize that stellar mass loss confined to the nuclear cusp or nuclear 
cluster does furnish a steady state, {\it in situ}\ supply of gas that is in 
principle available for accretion.  Shortly after being shed, the stellar 
gaseous envelopes quickly become thermalized with the hot ambient medium of
the bulge, but some of the gas remains cool (Parriott \& Bregman 2008).  
Even after the stellar ejecta joins the hot phase, the cooling time is 
sufficiently short in the inner region of the bulge that a cooling flow should 
develop (Mathews 1990).
 
The three main Local Group galaxies (M31, M32, and M33) serve as instructive 
examples.  From the work of Lauer et al. (1998), the central stellar densities 
of all three galaxies rise steeply toward the center as $\rho\,\propto\,
r^{-1.5\pm 0.5}$; at $r = 0.1$ pc, the density reaches 
$\rho\, \approx\, 10^{6.5\pm 0.3}$ \solmass\ pc$^{-3}$.  More typically, for 
galaxies beyond the Local Group, \hst\ data probe $r \approx 10$ pc, where 
$\rho\, \approx\, 10-10^3$ $L_{\odot, V}$ pc$^{-3}$ for the ``core'' 
ellipticals and $\rho\, \approx\, 10^2-10^4$ $L_{\odot, V}$ pc$^{-3}$ for the 
``power-law'' ellipticals and bulges of early-type spirals and S0s (e.g., 
Faber et al. 1997).   Within a spherical region of $r = 10$ pc, the diffuse 
cores have $L \approx 4\times 10^4 - 4\times 10^6\, L_{\odot, V}$, which 
yields $\dot M_*\,\approx\,1\times10^{-6} - 1\times10^{-4}$ \solmass\ 
\peryr; for the denser power-law cusps, $L \approx 4\times 10^5 - 
4\times 10^7\, L_{\odot, V}$, or $\dot M_*\,\approx\,1\times10^{-5} - 
1\times10^{-3}$ \solmass\ \peryr.  Centrally dominant nuclear star clusters, 
present in a large fraction of disk galaxies, typically have luminosities 
$L \approx 10^7$ \solum\ (Lauer et al. 1995; Carollo et al. 1997; B\"oker et 
al. 2002), and hence $\dot M_*\,\approx\,10^{-3}$ \solmass\ \peryr.

\subsection{Bondi Accretion}

The inner regions of ellipticals and bulges contain X-ray-emitting plasma, 
sustained through thermalized ejecta from stellar mass loss (Mathews 1990),
with temperatures characteristic of the virial velocities of the stars, 
$\sim 10^6-10^7$ K.  This diffuse, hot gas holds another plentiful fuel 
reservoir for accretion, through the mechanism described by Bondi (1952).  The 
Bondi accretion rate depends on the gas density and temperature at the 
accretion radius, $R_a \approx GM/c_s^2$, where $c_s \approx 0.1 T^{1/2}$ \kms\
is the sound speed of the gas.  From the continuity equation, 
$\dot M_{\rm B}\,=\,4\pi R_a^2 \rho_a c_s$, where $\rho_a$ is the gas density 
at $R_a$.   Expressed in terms of typical observed parameters (see below), 

\begin{equation}
\dot M_{\rm B}\,\approx\,7.3\times 10^{-4}\,
\left(\frac{M_{\rm BH}}{10^8\, M_{\odot}}\right)^2\, 
\left(\frac{n}{0.1\, {\rm cm}^{-3}}\right)\, 
\left(\frac{200\, {\rm km\,s}^{-1}}{c_s}\right)^3\,
\,M_{\odot}\,{\rm yr}^{-1}.
\end{equation}

Recent high-resolution X-ray observations using {\it Chandra}\ find that the 
diffuse gas in the central few hundred parsec regions of elliptical galaxies 
typically has temperatures of $kT\,\approx\,0.3-1$ keV and densities of 
$n\,\approx\,0.1-0.3$ \cc\ (Di~Matteo et al. 2001; Loewenstein et al. 2001; 
Sarazin et al. 2000; Pellegrini 2005; Soria et al. 2006).  Data for the bulges 
of spiral and S0 galaxies are more fragmentary.  {\it Chandra}\ has so far 
resolved the hot gas in the centers of a handful of bulges, including the 
Milky Way (Sbc; Baganoff et al. 2003), M81 (Sab; Swartz et al. 2003), NGC~1291 
(Sa; Irwin et al. 2002), and NGC~1553 (S0; Blanton et al. 2001).  The center 
of M31 (Sb) has been investigated using both {\it XMM-Newton}\ (Shirey et al. 
2001) and {\it Chandra}\ (Garcia et al. 2005).  These studies suggest that 
bulges typically have gas temperatures of $kT\,\approx\, 0.3-0.5$ keV.  
Information on gas densities is sketchier, but judging from the work on M81 
and NGC~1291, a fiducial value might be $n\,\approx\,0.1$ \cc.

If, for simplicity, we assume that the hot gas in most bulges is characterized 
by $n\,=\,0.1$ \cc\ and $kT\,=\,0.3$ keV, then $\dot M_{\rm B}\,\approx\,
10^{-5}-10^{-3}$ \solmass\ \peryr\ for \mbh\ = $10^7-10^8$ \solmass.  In 
elliptical galaxies \mbh\ $\approx\,10^8-10^9$ \solmass, and for $n\,=\,0.2$ 
\cc\ and $kT\,=\,0.5$ keV, $\dot M_{\rm B}\,\approx\, 10^{-4}-10^{-2}$ 
\solmass\ \peryr.  We note that these estimates of the Bondi accretion rates 
are probably lower limits because the actual densities near $R_a$ are
likely to be higher than we assumed.  For the above fiducial temperatures and 
BH masses, $R_a\,\approx\,1-10$ pc for bulges and $\sim 10-100$ pc 
for ellipticals, roughly an order of magnitude smaller than the typical linear 
resolution achieved by {\it Chandra}\ for nearby galaxies.  In the case of the 
Galactic center, $n = 26$ \cc\ and $kT = 1.3$ keV at 10\asec\ (0.4 pc), rising 
to $n = 130$ \cc\ and $kT = 2$ keV at 1\asec\ (Baganoff et al. 2003).

\subsection{Other Sources}

The two processes discussed above---ordinary stellar mass loss and Bondi 
accretion of hot gas---can be regarded as a conservative, steady state 
supply of fuel for galactic nuclei.   Other sources, however, can raise this 
minimum level.  In terms of purely stellar sources, some possibilities include 
(1) stellar mass loss enhanced by dynamical heating (Allen \& Hughes 1987; 
Armitage et al. 1996) or AGN irradiation (Edwards 1980; Scoville \& Norman 
1988; Voit \& Shull 1988), (2) stellar-stellar collisions in a dense nuclear 
cluster (Spitzer \& Saslaw 1966; Frank 1978; Rauch 1999), and (3) tidal 
disruption of stars by the central BH (Hills 1975; Rees 1988; Milosavljevi\'c 
et al. 2006). It is difficult to evaluate quantitatively the contribution 
these effects would make to the total fuel budget of nearby nuclei; we merely 
note that cumulatively they may significantly boost the ``baseline'' accretion 
rate estimated above.

We have also neglected any contribution from the cold phase of the interstellar
medium.  Nonaxisymmetric perturbations due to galaxy-galaxy tidal 
interactions, large-scale bars, nuclear bars, or nuclear spirals are often 
invoked as mechanisms for angular momentum transport of the cold gas in disk 
galaxies (e.g., Wada 2004).  The effectiveness of these processes for fueling 
nearby, relatively low-luminosity AGNs, however, has been unclear.  
With respect to the well-studied Palomar survey, AGN activity seems to be 
affected neither by large-scale bars (Ho et al. 1997c) nor by local galaxy 
environment (Schmitt 2001; Ho et al. 2003a).  In any event, if dissipation of 
the cold gas does occur on nuclear scales, as inevitably it must at some 
level in at least some objects, it would further add to the fuel supply.

\section{Implications}

\subsection{Accretion Flow and Radiative Efficiency}

The results presented in this paper provide some important insights into the 
nature of BH accretion in nearby galactic nuclei.  We have established, for 
the first time using a large, statistically robust sample, that virtually 
{\it all}\ massive BHs in the nearby Universe share two common properties: 
they have low luminosities and radiate well below the Eddington limit.  This 
holds for galaxies spanning a wide range of Hubble types and nuclear spectral 
classes.  The median value of the bolometric luminosities are only \lbol\ 
$\approx\, 10^{39} - 10^{41}$ \lum, and the median Eddington ratios range from 
\lbol/\ledd\ $\approx\,2\times 10^{-7}$ to $3\times 10^{-4}$.  

The extreme dimness of these nuclei strongly suggests that their accretion 
flows are radiatively inefficient.  In the context of the class of accretion 
models commonly called optically thin RIAFs\footnote{As originally formulated
by Narayan \& Yi (1994, 1995), this class of accretion disk models was called
advection-dominated accretion flows.  Subsequent work has shown that 
such flows are inherently unstable to outflows and convection.  To avoid 
delving into the technical details, which are unimportant for the present 
level of discussion, we simply follow Quataert (2001) and refer to this class
of models as RIAFs.} the accretion luminosity is given by (Mahadevan 1997)
$L_{\rm acc}=\left(\eta/0.1\right)\left[0.20(\dot m/\alpha^2)\right]\dot Mc^2$,
valid in the regime $\dot m > 10^{-3} \alpha^2$, where 
$\dot M = \dot m \dot M_{\rm Edd}$ and 
$\dot M_{\rm Edd} = 2.2\times 10^{-8} \left(\eta/0.1\right) 
\left(M_{\rm BH}/M_{\odot}\right)$ \solmass\ \peryr. This expression adopts the 
canonical values of the microphysics parameters used by Mahadevan (1997).  In 
the notation used in this paper, 

\begin{equation}
\dot m\,\simeq \, 0.7\,\left(\frac{\alpha}{0.3}\right)\,\left(\frac{L_{\rm bol}}{L_{\rm Edd}}\right)^{1/2}.
\end{equation}

\noindent
Thus, for \lbol/\ledd\ = $10^{-6} - 10^{-4}$, $\dot m \approx 3\times 10^{-4} 
- 2\times 10^{-2}$, which lie comfortably within the regime of optically thin 
RIAFs, $\dot m \leq \dot m_{\rm crit} \approx \alpha^2 \approx 0.1$ (Narayan 
et al. 1998).  The corresponding absolute mass accretion rates, for \mbh\ 
$\approx\, 10^6 - 10^9$ \solmass, are $\dot M \approx 10^{-5} - 10^{-1}$ 
\solmass\ \peryr.  

The accretion rates estimated in \S~4 provide a more direct argument that the 
radiative efficiency of the accretion flow, whatever its form, is likely to be 
low.  The median bolometric luminosities of the Palomar emission-line objects 
range from $\sim$1\e{40} \lum\ for transition objects to 
$\sim$2\e{41} \lum\ for Seyferts (Table~2).  If this emission is produced by a 
canonical optically thick, physically thin disk, which radiates at  
$L_{\rm acc} = \eta \dot M c^2 = 5.7\times10^{45} \left(\eta/0.1\right) 
\left(\dot M/M_{\odot}\, {\rm yr}^{-1}\right)$ \lum, we expect typical mass 
accretion rates of $\dot M\, \approx\,2\times 10^{-6}$ to 4\e{-5} \solmass\ 
\peryr.  These values of $\dot M$ are miniscule by comparison with the minimum 
accretion rates likely to be available through stellar mass loss and Bondi 
accretion alone.  The majority of the Palomar AGNs are hosted by early-type 
disk galaxies (S0s and Sa--Sbc spirals; see Ho et al. 1997b, 2003a), whose 
bulges tend to have cuspy light profiles of the ``power-law'' type.  As 
discussed in \S~4.1, the inner regions of such bulges should have 
$\dot M_*\,\approx\, 10^{-5}-10^{-3}$ \solmass\ \peryr, and probably closer to 
the upper end of this range because of the additional contribution from 
central star clusters, which add $\dot M_*\, \approx\, 10^{-3}$ \solmass\ 
\peryr.  We have also erred on the side of caution by assuming that all of the 
stars are old.  In actuality, the central regions of many spirals in the 
Palomar sample often show evidence for some contribution from composite 
populations (Ho et al. 2009), which will help to boost the mass loss rates 
even further.  Bondi accretion of hot gas contributes roughly the same amount 
as stellar mass loss, $\dot M_{\rm B}\,\approx\, 10^{-5}-10^{-3}$ \solmass\ 
\peryr.  Thus, $\dot M_{\rm tot} = \dot M_* + \dot M_{\rm B}\,\approx\, 10^{-5}
-10^{-3}$ \solmass\ \peryr, but more likely, $\dot M_{\rm tot}$ \gax\ $10^{-3}$
\solmass\ \peryr.  

A similar exercise leads to an even stronger result for the absorption-line 
nuclei, which are found predominantly in elliptical and S0 galaxies.  Here, 
the median \lbol\ of 2\e{39} \lum\ requires only $\dot M \approx 4\times10^{-7}$
\solmass\ \peryr\ for $\eta$ = 0.1.  On the other hand, the centers of the 
host galaxies can supply at least $\dot M_*\,\approx\,10^{-6} - 10^{-4}$ 
\solmass\ \peryr\ for low-density cores, a factor of 10 higher still in 
$\dot M_*$ for power-law cusps, and yet another 10-fold increase for 
$\dot M_{\rm B}$.  Ho et al. (2003b) highlighted the acuteness of the 
luminosity-deficit problem for the nearest of the absorption-line objects, 
M32, whose 2.5\e{6} \solmass\ BH emits merely 9.4\e{35} \lum\ in the 2--10 keV 
band at an Eddington ratio of $\sim 3\times 10^{-9}$.  

These simple comparisons suggest that if $\eta$ indeed is 0.1, then nearby 
galactic nuclei are 1--4 orders of magnitude underluminous.  For the accretion 
rates that we infer to be present, they should radiate far more prodigiously 
than actually observed.  There are four possible interpretations of this 
finding.  (1) First, our estimates of $\dot M_{\rm tot}$ could be too high by 
a large factor, namely 1--4 orders of magnitude.  We consider this to be 
unlikely.  Recall that $\dot M_{\rm tot}$ includes only normal mass loss from 
evolved stars and Bondi accretion of hot gas, either one of which alone would 
violate the luminosity limit for $\eta$ = 0.1; we have conservatively 
neglected other potential sources of fuel (\S~4.3).  (2) Second, it could be 
argued that perhaps the gas released through stellar mass loss manages to 
escape from the nucleus before it gets accreted.  In actively star-forming 
galaxies, for example, the collective effects of strong winds and shocks from 
massive stars can expel gas to large galactocentric distances.  This mechanism 
of gas removal, however, appears to be extremely implausible given the typical 
ages of the nuclear stellar population (Ho et al. 2003a).  Storing the 
gas in an inert cold disk or converting it to young stars violates other 
observational constraints (Ho 2008). (3) Third, $\eta$ may
be $\ll 0.1$, as expected from RIAFs (Narayan et al. 1998; Quataert 2001). This
is an argument that is frequently invoked to explain the apparent conflict 
between the nuclear luminosities and Bondi accretion rates in some early-type
galaxies (e.g., Fabian \& Rees 1995; Mahadevan 1997; Di~Matteo et al.  2000, 
2001; Loewenstein et al. 2001; Ho et al. 2003b).  (4) Lastly, an inherent 
ambiguity exists between inefficient radiation and inefficient accretion.  
A variety of physical effects, summarized in Ho (2008), can divert the 
inflowing gas and severely curtail the amount of material that ultimately gets 
accreted.  For instance, if RIAFs naturally develop outflows or winds, as 
recently shown in a number of studies (see Quataert 2001), the actual 
accretion rate would be much lower than that estimated at large radii.  If so, 
then $\eta$ may not need to be so exceptionally low, although it should still 
be substantially below 0.1 because the outflow models ultimately rely on the 
accreting gas to be radiatively inefficient.  Energy feedback from the AGN, 
associated with either disk outflows or small-scale radio jets---a ubiquitous 
feature of low-\lbol/\ledd\ systems (Ho 2002a)---may be another culprit for 
interrupting smooth mass inflow.  Ho (2009) presents quantitative evidence 
that AGN feedback injects nongravitational perturbations to the kinematics of 
the ionized gas in the Palomar sources.

\subsection{Accretion States of Massive Black Holes}

By analogy with stellar BHs in X-ray binaries, supermassive BHs in galactic 
nuclei may evolve through different evolutionary phases, corresponding to 
distinct ``states'' (Narayan et al. 1998).  The basic physical picture is that 
the structure of the accretion flow changes in response to variations in the 
accretion rate.  Parameterizing the accretion rate in terms of the 
dimensionless variable $\dot m$, one can define three, possibly four regimes.  
(1) When $\dot m$ \gax\ 1, an object is in the ``very high'' state.  The 
high radiation density in these ``super-Eddington'' sources traps the 
photons, and the accretion flow is that of an optically thick RIAF or slim 
disk (Begelman \& Meier 1982; Abramowicz et al. 1988). 
An extragalactic analog of such systems are the narrow-line Seyfert 1 
nuclei (Pounds \& Vaughan 2000).  (2) Objects satisfying $\dot m_{\rm crit} 
< \dot m < 1$ correspond to those in the ``high'' state.   These contain 
traditionally studied optically thick, geometrically thin, radiatively 
efficient Shakura \& Sunyaev (1973) disks, which are present in classical, 
relatively luminous Seyfert nuclei and quasars.  (3) When $\dot m \leq \dot 
m_{\rm crit}$, an optically thin, geometrically thick, radiatively inefficient 
flow develops, and the luminosity of the source plummets.  Depending on how 
low the accretion rate drops, one might distinguish between objects in the 
``low'' state ($10^{-6} \leq \dot m \leq \dot m_{\rm crit}$) versus those in 
the ``quiescent'' state ($\dot m < 10^{6}$).  Objects in quiescence are those 
dominated by, or which exclusively contain, a pure RIAF, such as the Galactic 
center source Sgr~A$^*$.  Low-state objects contain a hybrid 
structure consisting of an inner RIAF plus an outer thin disk, whose 
truncation radius recedes as $\dot m$ decreases.  Such a configuration
has been suggested for a number of low-luminosity AGNs, especially LINERs 
(Lasota et al. 1996; Quataert et al. 1999; Ho et al. 2000; Ho 2002b, 2008).
We note that the boundary between the low and quiescent states ($\dot m 
\approx 10^{-6}$) is purely illustrative; it remains to be demonstrated 
that there are two distinct states, and if so, where the transition truly lies.

If AGN activity is characterized by distinct states, as schematically sketched 
above, then this ought to be reflected in the observed distribution of 
accretion rates for AGNs spanning the full range of $\dot m$.  Since it is 
difficult to measure $\dot m$ directly, \lbol/\ledd\ can be used as a surrogate 
for $\dot m$.  We stress, however, that analysis of this kind is only 
meaningful when performed on large, well-defined, statistically complete 
samples.  It is dangerous to combine samples with different selection criteria 
(e.g., Marchesini et al. 2004; Hopkins et al. 2006).  With this in mind, we 
recall that the distribution of \lbol/\ledd\ for the entire Palomar sample 
(Figs.~1{\it b}\ and 3{\it b})---a more or less complete census of nearby 
galaxies---shows no obvious substructure that might be identified with 
physically distinct populations.  Unfortunately, the volume sampled by the 
Palomar survey does not contain sufficient luminous AGNs to properly cover the 
upper end of the \lbol/\ledd\ distribution.  This would be an important goal 
for future statistical studies of AGNs.  

\section{Summary}

Nearly all of the objects in the Palomar survey of nearby galaxies now have 
central stellar velocity dispersions, from which BH masses can be inferred
using the \mbh$-$$\sigma_*$ relation.  A previously published collection of 
H\al\ line fluxes, in combination with a newly assembled database of nuclear 
X-ray measurements and a reevaluation of the appropriate bolometric 
corrections, provides estimates of the accretion luminosity of the nuclei.  
We use these resources to evaluate the distribution of bolometric luminosities 
and Eddington ratios for a large, well-defined sample of galactic nuclei in 
order to investigate the nature of accretion onto massive BHs in nearby 
galaxies.  

Nearby galactic nuclei span at least 7 orders of magnitude in nuclear 
bolometric luminosities, from \lbol\ $<\, 10^{37}$ to $\sim 3\times 10^{44}$ 
\lum, and an even broader range in Eddington ratios, from \lbol/\ledd\ 
$\approx\, 10^{-9}$ to $10^{-1}$.   Both \lbol\ and \lbol/\ledd, but 
especially the latter, decrease systematically along the following 
spectral sequence: Seyferts $\rightarrow$ LINERs $\rightarrow$ transition 
objects $\rightarrow$ absorption-line nuclei.  The spectral diversity of 
emission-line nuclei reflects and is primarily controlled by variations in the 
mass accretion rate.  The characteristic value of \lbol/\ledd\ also varies 
systematically along the Hubble sequence, increasing from galaxies with large 
to small bulge-to-disk ratios.

The accretion rates inferred from the nuclear luminosities, assuming a standard
radiative efficiency of $\eta = 0.1$, are very low, typically $\dot M$
\lax\ $10^{-6}$ to $10^{-5}$ \solmass\ \peryr.  Such tiny rates can easily be 
supplied {\it in situ}\ by ordinary mass loss from evolved stars in the 
nuclear stellar cusp or central star clusters, or by Bondi accretion of hot 
gas in the inner bulges of galaxies detected in X-ray observations.  Indeed, we 
argue that conservative estimates of the gas mass potentially available
for accretion already far exceed the observational limits imposed by 
the luminosity measurements, suggesting that in many, if not most, nearby 
galaxies the radiative efficiency is likely to be much less than 0.1.  The 
requirement for exceptionally low radiative efficiencies, however, could be 
mitigated if the actual accretion rates are curtailed by AGN feedback in the 
form of disk outflows or small-scale radio jets.  The prevalence of 
RIAFs is further supported by the low 
Eddington ratios: all the objects in our sample are sub-Eddington systems, 
with the majority having values of \lbol/\ledd\ that satisfy the theoretically 
predicted criterion for RIAFs.

We suggest that massive BHs in galactic nuclei evolve through distinct states 
in response to changes in the mass accretion rate.  The nearby objects 
considered in this study are largely systems in the low or quiescent state.  We
see no evidence of bimodality in the distribution of \lbol/\ledd, but this is 
probably a consequence of the limited volume probed by the Palomar survey.  It 
would be of considerable interest to extend the analysis presented here to 
include objects of higher luminosity in order to map out the full distribution 
of AGN luminosities and Eddington ratios. 

\acknowledgements
This work was supported by the Carnegie Institution of Washington and by {\it 
Chandra}\ grant GO5-6107X.  I thank Louis-Benoit Desroches for help with 
analyzing some of the archival {\it Chandra}\ data.  An anonymous 
referee offered a positive and helpful report.


\clearpage
\appendix

\section{Nuclear X-ray Luminosities}

This Appendix summarizes the X-ray luminosities used in this paper.  We 
performed a comprehensive literature search of all published X-ray measurements
for the AGNs (Seyferts, LINERs, and transition objects) and absorption-line
nuclei in the Palomar survey.  Because of the faintness of most nearby nuclei 
and potential confusion with circumnuclear emission, the most crucial 
consideration is angular resolution.  With a point-spread function (PSF) of 
FWHM $\approx$ 1\asec\ and low background noise, the instrument of choice is 
ACIS on {\it Chandra}.  Although less ideal, the HRI imager on {\it ROSAT}\ 
and the MOS camera on {\it XMM-Newton}, both having PSFs with FWHM $\approx$
5\asec, also yield acceptable data under most circumstances.  For sources 
bright enough for rigorous spectral analysis, even lower resolution 
observations (e.g., {\it ASCA}) can be used if the nucleus can be isolated
through spectral fitting.

Of the 277 galaxies in the parent sample, acceptable literature data were 
located for 166.  Most of the observations (75\%) were acquired with 
{\it Chandra}/ACIS, and the rest were taken largely with {\it ROSAT}/HRI or 
{\it XMM-Newton}.  Only a small handful come from {\it ROSAT}/PSPC and 
{\it ASCA}; one object was observed with {\it BeppoSAX}.  We also performed a
thorough search of the {\it Chandra}\ archives and included all useful, 
unpublished data that were nonproprietary as of 2007 October.  A total of nine 
additional galaxies were located.  We analyzed these data sets following 
standard techniques, as described in Ho et al. (2001) and Desroches \& Ho 
(2009).

Table~3 lists X-ray luminosities for the final sample of 175 galaxies. Because 
the literature data were acquired with a variety of different instruments and 
analyzed using many different techniques, we converted all the luminosities 
to one standard bandpass, 2--10 keV.  When reliable spectral fits are 
available, we use the published best-fit spectral slope to extrapolate to the 
desired bandpass.  Otherwise, we assume a photon index of $\Gamma = 1.8$, which 
is close to the typical values observed in low-luminosity AGNs (see Ho 2008, 
and references therein).  Likewise, we quote intrinsic (absorption-corrected) 
luminosities whenever possible.  Although many of the fainter sources do not 
have sufficient counts to constrain the absorbing column, many lines of 
evidence suggest that most low-luminosity AGNs do not suffer from much 
obscuration (Ho 2008).

\clearpage
\begin{figure*}[t]
\psfig{file=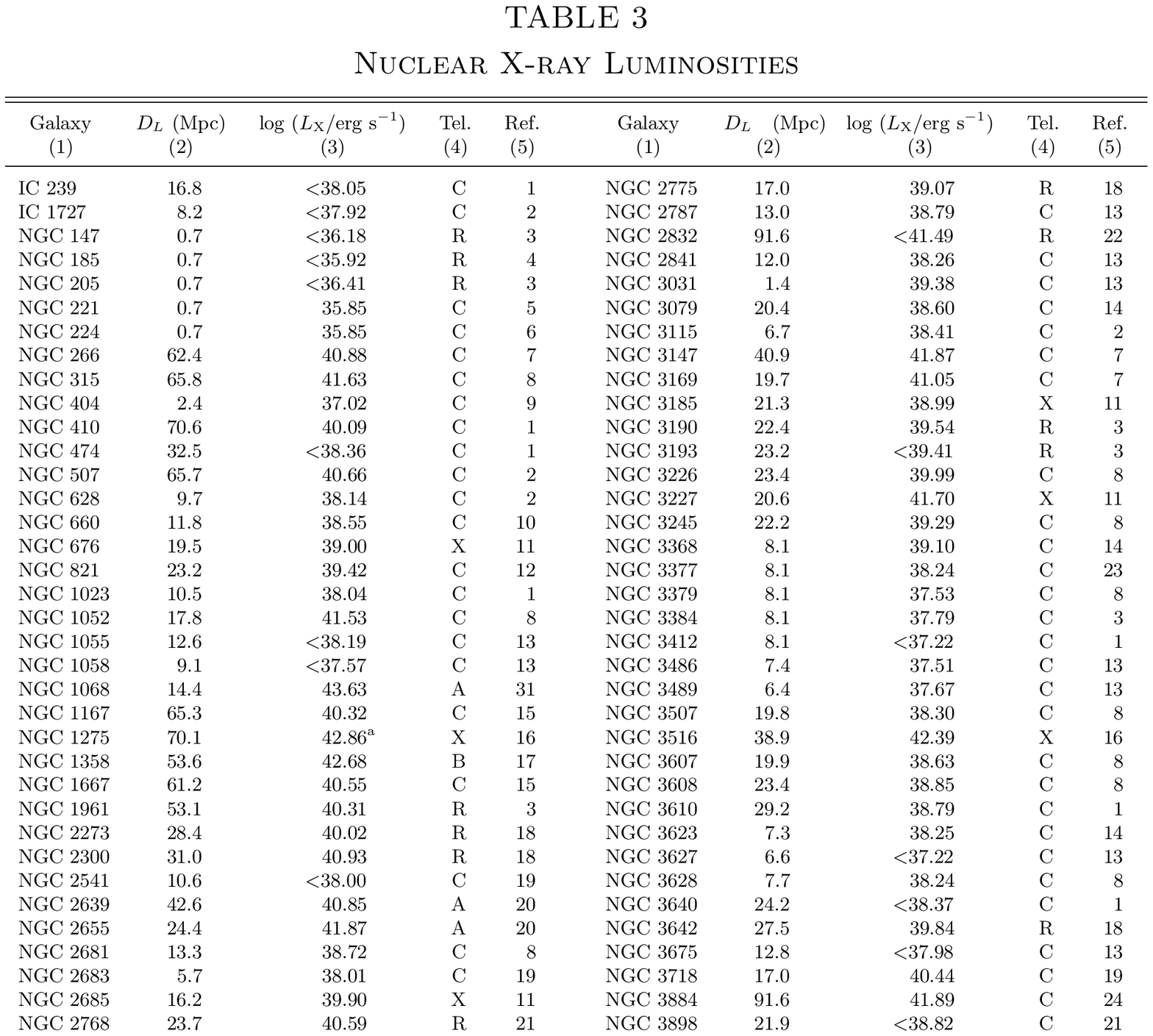,width=18.5cm,angle=0}
\end{figure*}
\clearpage
\begin{figure*}[t]
\psfig{file=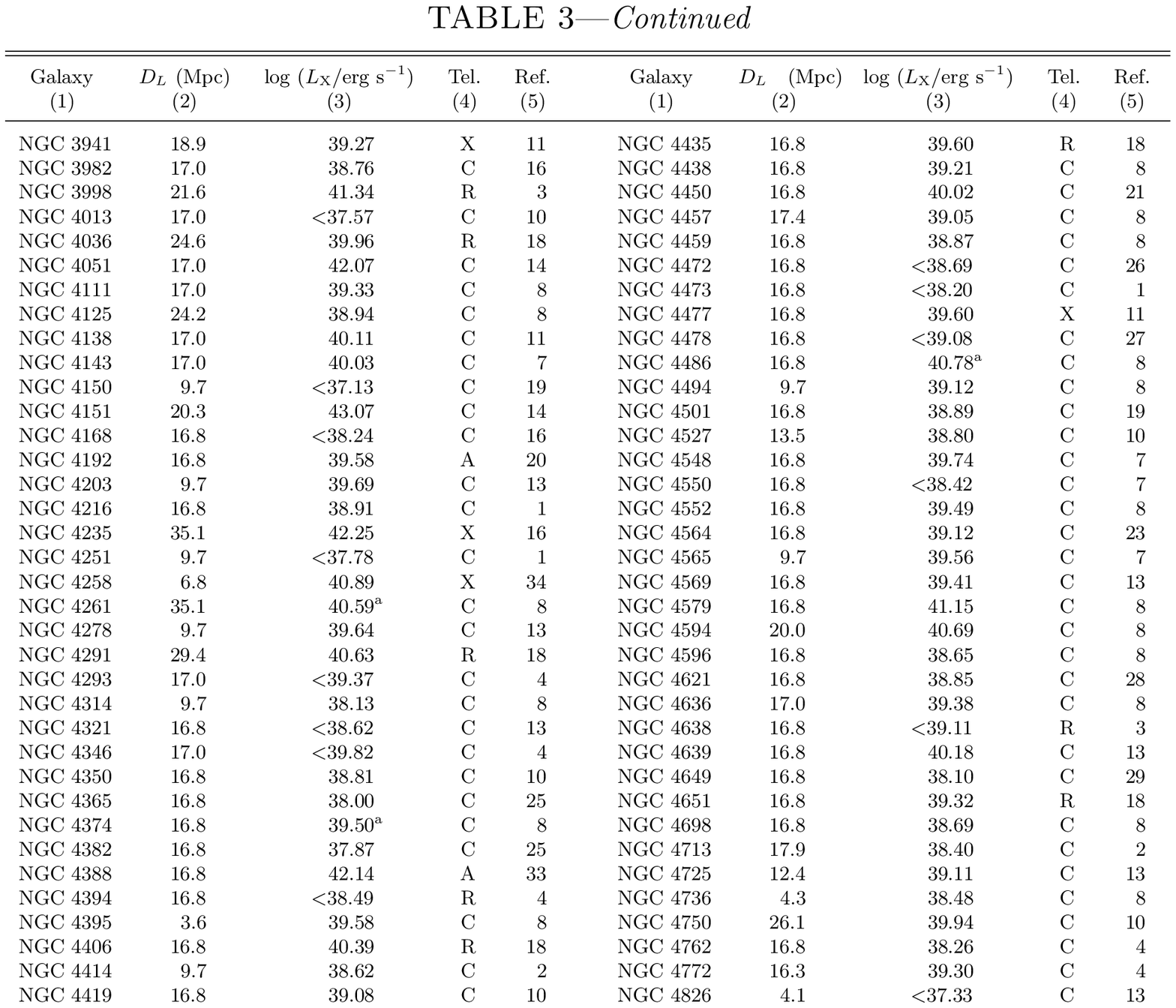,width=18.5cm,angle=0}
\end{figure*}
\clearpage
\begin{figure*}[t]
\psfig{file=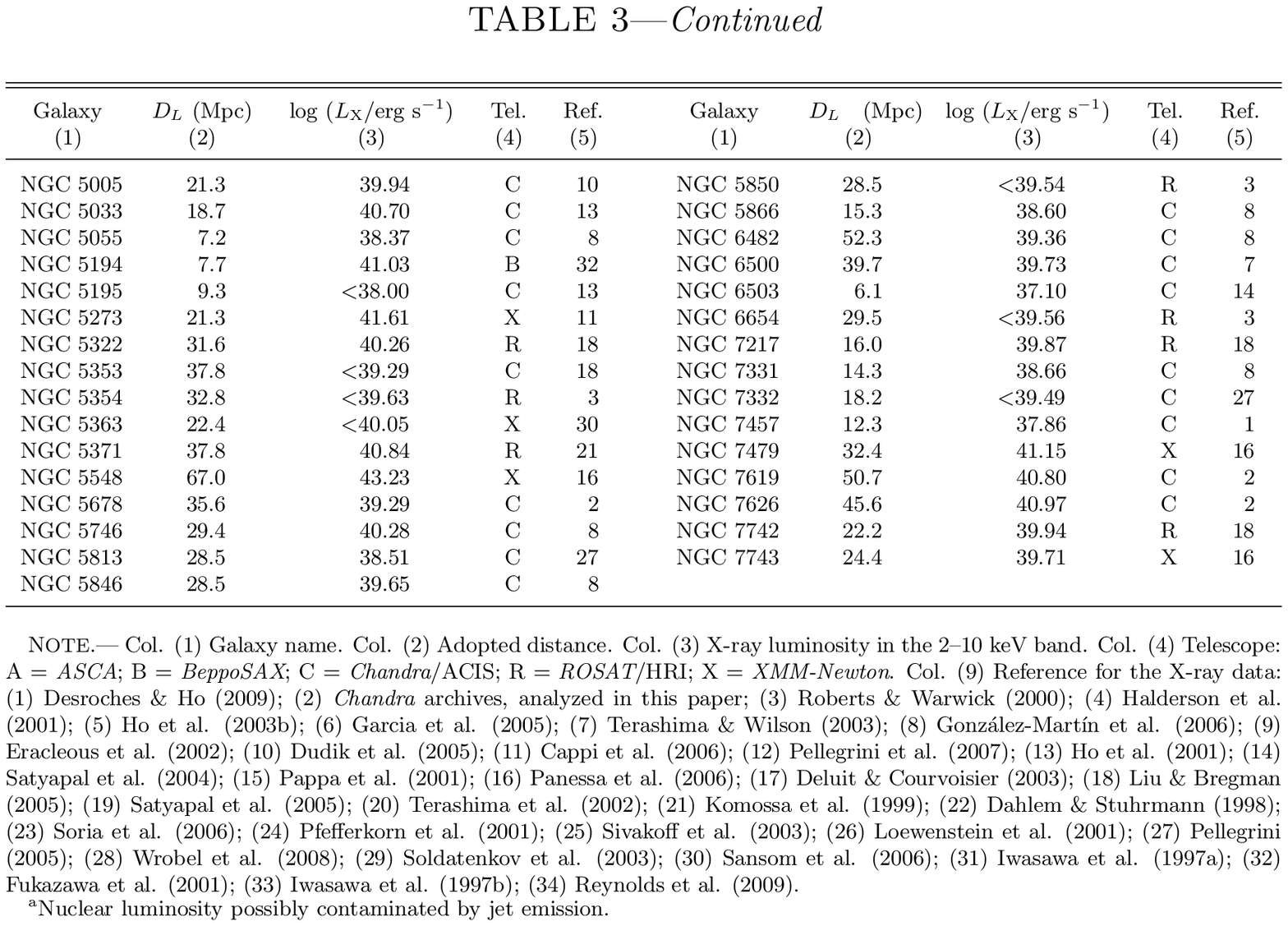,width=18.5cm,angle=0}
\end{figure*}

\end{document}